\documentclass[useAMS,usenatbib]{mn2e}
\usepackage{epsf,graphicx,rotating,url}

\title[The BELR outer radius]{Constraints on the outer radius of the broad emission line region of active galactic nuclei}

\author[H. Landt et al.]{Hermine Landt$^1$\thanks{E-mail:
    hermine.landt@durham.ac.uk}\thanks{Visiting Astronomer at the
    Infrared Telescope Facility, which is operated by the University
    of Hawaii under Cooperative Agreement no. NNX-08AE38A with the
    National Aeronautics and Sp ace Administration, Science Mission
    Directorate, Planetary Astronomy Program.}, Martin
  J. Ward$^1$\footnotemark[2], Martin Elvis$^2$\footnotemark[2] and Margarita Karovska$^2$ \\ 
$^1$Department of Physics, Durham University, South Road, Durham, DH1 3LE \\ 
$^2$Harvard-Smithsonian Center for Astrophysics, 60 Garden Street, 
Cambridge, MA 02138, USA}

\begin{document}

\def\la{\mathrel{\hbox{\rlap{\hbox{\lower4pt\hbox{$\sim$}}}\hbox{$<$}}}}
\def\ga{\mathrel{\hbox{\rlap{\hbox{\lower4pt\hbox{$\sim$}}}\hbox{$>$}}}}

\font\sevenrm=cmr7
\def\OI{O~{\sevenrm I}}
\def\CaII{Ca~{\sevenrm II}}
\def\FeII{Fe~{\sevenrm II}}
\def\OIII{[O~{\sevenrm III}]}
\def\FeIIf{[Fe~{\sevenrm II}]}
\def\OII{[O~{\sevenrm II}]}
\def\SIII{[S~{\sevenrm III}]}

\def\cloudy{{\sevenrm CLOUDY}}

\date{Accepted ~~. Received ~~; in original form ~~}

\pagerange{\pageref{firstpage}--\pageref{lastpage}} \pubyear{2014}

\maketitle

\label{firstpage}

\begin{abstract}

Here we present observational evidence that the broad emission line
region (BELR) of active galactic nuclei (AGN) generally has an outer
boundary. This was already clear for sources with an obvious
transition between the broad and narrow components of their emission
lines. We show that the narrow component of the higher-order Paschen
lines is absent in {\it all} sources, revealing a broad emission line
profile with a broad, flat top. This indicates that the BELR is
kinematically separate from the narrow emission line region. We use
the virial theorem to estimate the BELR outer radius from the flat top
width of the unblended profiles of the strongest Paschen lines,
Pa$\alpha$ and Pa$\beta$, and find that it scales with the ionising
continuum luminosity roughly as expected from photoionisation
theory. The value of the incident continuum photon flux resulting from
this relationship corresponds to that required for dust sublimation. A
flat-topped broad emission line profile is produced by both a
spherical gas distribution in orbital motion as well as an accretion
disc wind if the ratio between the BELR outer and inner radius is
assumed to be less than $\sim 100 - 200$. On the other hand, a pure
Keplerian disc can be largely excluded, since for most orientations
and radial extents of the disc the emission line profile is
double-horned.

\end{abstract}


\section{Introduction}

The broad emission line region (BELR) of active galactic nuclei (AGN)
is one of the most direct tracers of the immediate environment of
supermassive black holes. However, despite decades of intensive
optical and ultraviolet (UV) spectrophotometric studies its geometry
and kinematics remain ill-defined. It is not clear whether the BELR
gas has a spherical, bi-conical or disc-like distribution, and whether
it consists of a large number of discrete clouds
\citep[e.g.][]{Math74, Kro81, Emm92, Bal95} or is part of an
outflowing, continuous gas distribution such as an accretion disc wind
\citep[e.g.][]{Murray95, Murray97}. The knowledge of the BELR geometry
and kinematics is essential not only to our understanding of the
relationship between different types of AGN but in particular to
studies that estimate black hole masses from the widths of the broad
emission lines. These estimates would not be meaningful if the BELR
gas kinematics were dominated by radial motion, and if the gas was
gravitationally bound but distributed in a disc, the resulting values
would be underestimated for sources viewed face-on.

Our current, limited knowledge of the BELR physical conditions and
scales was gained primarily through the application of photoionisation
models to the observed emission line intensities and ratios
\citep[see, e.g., review by][]{Ferl03} and through reverberation
mapping studies of the (correlated) continuum and line variability
\citep[see, e.g., review by][]{Pet93}. Both of these methods have
relied so far on spectroscopic observations at optical and UV
frequencies, but investigating also the near-infrared (near-IR)
wavelengths can offer several advantages. First, the near-IR broad
emission lines are little reddened and, therefore, when compared to
lines at higher frequencies, can yield information on the amount of
dust extinction affecting the BELR. Secondly, since each emission line
is formed most efficiently at a particular density and distance from
the ionising source \citep{Bal95}, it is important to incorporate in
photoionisation and reverberation mapping studies multiple emission
lines to map the entire BELR. In this respect, cross-dispersed near-IR
spectra, which are now available at several observing sites, have a
large wavelength coverage and so offer the unique opportunity to
simultaneously observe a plethora of different emission lines. For
example, in low-redshift sources, such spectra cover the wavelength
region of $\sim 16$ hydrogen lines from the Paschen and Brackett
series, as well as that of several lines from other chemical species
such as helium, oxygen, calcium and singly-ionised iron \citep{L08a}.

Finally, the near-IR broad emission lines trace in depth the
low-ionisation line (LIL) region, a region believed to have extreme
properties such as very high densities (\mbox{$n>10^{11}$ cm$^{-3}$})
and a disc-like structure \citep{Coll88, Marz96}. However, these
properties were derived mainly based on observations of the two Balmer
lines, H$\alpha$ and H$\beta$, which are strongly blended with other
species. Since the crucial measurements for these studies are the
emission line flux and profile a verification of the results using
{\it unblended} broad emission lines is needed. As we have shown in
\citet{L08a}, the profiles of the strongest Paschen lines, Pa$\alpha$
and Pa$\beta$, are observed to be unblended and, therefore, are
well-suited to test BELR models. However, in order to do so in a
meaningful way one needs to first isolate the profile of the broad
component from that of the narrow emission line region (NELR), which
is not a straightforward task in sources where the broad emission
lines are relatively narrow and so the narrow and broad components
smoothly merge into each other.

Here we present observational evidence that the BELR of {\it all} AGN
has an outer boundary, which means that it is kinematically separate
from the NELR. The paper is organised as follows. In Section 2, we
briefly introduce the sample and discuss the data. In Section 3, based
on the higher-order Paschen lines, we show that the intrinsic BELR
profile has a broad, flat top, which indicates an outer radius. In
Section 4, we estimate the BELR outer radius from the flat top width
of the unblended profiles of the strongest Paschen lines, Pa$\alpha$
and Pa$\beta$, using the virial theorem and investigate if the BELR is
dust-limited. In Section 5, we discuss BELR models in the light of
their ability to produce a flat-topped profile. Finally, in Section 6,
we summarise our main results and present our conclusions. Throughout
we have assumed cosmological parameters $H_0 = 70$ km s$^{-1}$
Mpc$^{-1}$, $\Omega_{\rm M}=0.3$, and $\Omega_{\Lambda}=0.7$.

\section{The sample and data}


\begin{table*}
\caption{\label{sample} 
Parameters of the Paschen broad emission line region}
\begin{tabular}{lclcrccrcrccr}
\hline
Object Name & z & line & type & top width & $M_{\rm BH}$ & Ref. & $R_{\rm H\beta}$ & Ref. & $R_{\rm out}$ & $\log \nu L_{\rm tot}$ & 
$T_{\rm dust}$ & $R_{\rm dust}$ \\
&&&& [km/s] & [$M_{\odot}$] && [lt-days] && [lt-days] & [erg s$^{-1}$] & [K] & [lt-days] \\
(1) & (2) & (3) & (4) & (5) & (6) & (7) & (8) & (9) & (10) & (11) & (12) & (13) \\
\hline
IRAS 1750$+$508 & 0.300 & Pa$\beta$  & e    &  884 & 3.3e$+$08 & est & 109 & est &  8604 & 47.07 & 1383 &  6556 \\
H 1821$+$643    & 0.297 & Pa$\beta$  & i    & 2050 & 1.9e$+$09 & est & 253 & est &  9487 & 47.22 & 1320 &  8796 \\
PDS 456         & 0.184 & Pa$\beta$  & e    &  998 & 8.0e$+$08 & est & 231 & est & 16382 & 47.41 & 1425 &  8972 \\
3C 273          & 0.158 & Pa$\beta$  & e    & 1080 & 8.9e$+$08 & P04 & 307 & B09 & 15551 & 47.59 & 1443 & 10683 \\
PG 0052$+$251   & 0.155 & Pa$\alpha$ & i    & 1523 & 3.7e$+$08 & P04 &  90 & B09 &  3260 & 46.07 & 1198 &  3012 \\   
PG 1307$+$085   & 0.155 & Pa$\alpha$ & i    &  729 & 4.4e$+$08 & P04 & 106 & B09 & 16950 & 46.22 & 1307 &  2854 \\
PG 0026$+$129   &0.145$^\star$&Pa$\alpha$&e &  684 & 3.9e$+$08 & P04 & 111 & B09 & 17199 & 46.41 & 1127 &  5221 \\
Mrk 876         & 0.129 & Pa$\alpha$ & i    &  877 & 2.8e$+$08 & P04 &  40 & B09 &  7441 & 46.22 & 1339 &  2680 \\
HE 1228$+$013   & 0.117 & Pa$\alpha$ & e    &  865 & 1.0e$+$08 & est &  53 & est &  2829 & 46.67 & 1333 &  4552 \\
PG 0804$+$761   & 0.100 & Pa$\alpha$ & lack &  505 & 6.9e$+$08 & P04 & 147 & B09 & 55723 & 45.97 & 1314 &  2111 \\
PG 1211$+$143   & 0.081 & Pa$\alpha$ & e    &  764 & 1.5e$+$08 & P04 &  94 & B09 &  5122 & 46.22 & 1337 &  2691 \\
PG 0844$+$349   & 0.064 & Pa$\alpha$ & lack &  651 & 9.2e$+$07 & P04 &  32 & B09 &  4472 & 46.18 & 1190 &  3478 \\
Mrk 1513        & 0.063 & Pa$\alpha$ & e    &  801 & 4.6e$+$07 & G12 &  10 & G12 &  1470 & 46.33 & 1356 &  2944 \\
3C 390.3        & 0.056 & Pa$\alpha$ & i    & 1478 & 1.3e$+$09 & D12 &  44 & D12 & 11810 & 44.95 & 1462 &   494 \\
Mrk 110         & 0.035 & Pa$\beta$  & e    &  747 & 2.5e$+$07 & P04 &  26 & B09 &   923 & 46.00 & 1452 &  1685 \\
Mrk 509         & 0.034 & Pa$\beta$  & e    & 1125 & 1.4e$+$08 & P04 &  80 & B09 &  2314 & 45.77 & 1398 &  1427 \\
Ark 120         & 0.033 & Pa$\beta$  & lack & 2546 & 1.5e$+$08 & P04 &  40 & B09 &   474 & 45.45 & 1102 &  1833 \\
3C 120          & 0.033 & Pa$\beta$  & i    &  679 & 6.7e$+$07 & G12 &  26 & G12 &  2978 & 45.28 & 1389 &   826 \\
Mrk 817         & 0.031 & Pa$\beta$  & i    & 1610 & 4.3e$+$07 & D10 &  14 & D10 &   342 & 45.65 & 1401 &  1236 \\
Mrk 290         & 0.030 & Pa$\beta$  & i    & 1426 & 2.4e$+$07 & D10 &   9 & D10 &   245 & 45.36 & 1353 &   969 \\
H 2106$-$099    & 0.027 & Pa$\beta$  & e    &  969 & 4.4e$+$07 & est &  11 & est &   953 & 45.28 & 1342 &   903 \\
Mrk 335         & 0.026 & Pa$\beta$  & e    &  945 & 2.5e$+$07 & G12 &  14 & G12 &   574 & 45.63 & 1308 &  1444 \\
Ark 564         & 0.025 & Pa$\beta$  & e    &  764 & 1.7e$+$07 & est &   8 & est &   614 & 45.37 & 1202 &  1334 \\
Mrk 79          & 0.022 & Pa$\beta$  & i    &  836 & 5.2e$+$07 & P04 &  15 & B09 &  1535 & 44.82 & 1364 &   510 \\
NGC 5548        & 0.017 & Pa$\beta$  & i    & 3731 & 4.4e$+$07 & D10 &  12 & D10 &    65 & 44.58 & 1547 &   279 \\
NGC 7469        & 0.016 & Pa$\beta$  & i    &  567 & 1.2e$+$07 & P04 &   5 & B09 &   777 & 45.28 & 1551 &   620 \\
H 1934$-$063    & 0.011 & Pa$\beta$  & e    &  745 & 1.0e$+$07 & est &   5 & est &   370 & 45.07 & 1426 &   605 \\
NGC 4593        & 0.009 & Pa$\beta$  & i    & 2777 & 9.8e$+$06 & D06 &   4 & B09 &    26 & 44.72 & 1380 &   441 \\
NGC 3516        & 0.009 & Pa$\beta$  & lack & 2728 & 3.2e$+$07 & D10 &  12 & D10 &    87 & ...   & ...  & ...   \\
NGC 3227        & 0.004 & Pa$\beta$  & i    & 1046 & 7.6e$+$06 & D10 &   4 & D10 &   143 & ...   & ...  & ...   \\
NGC 4151        & 0.003 & Pa$\beta$  & i    & 2272 & 4.6e$+$07 & B06 &   7 & B09 &   181 & 43.24 & 1328 &    89 \\
\hline
\end{tabular}

\parbox[]{17.5cm}{The columns are: (1) object name; (2) redshift from
  the NASA/IPAC Extragalactic Database (NED), which we checked with
  narrow emission lines; (3) emission line used as a template profile;
  (4) type of transition between the broad and narrow emission line
  component, where i: inflected (i.e., transition is obvious), e:
  estimated (i.e., transition is estimated), and lack: absent narrow
  emission line component; (5) full width of the flat top of the broad
  component; (6) black hole mass (in solar masses); (7) reference for
  the black hole mass, where B06: \citet{Bentz06b}, D06:
  \citet{Denney06}, D10: \citet{Denney10}, D12: \citet{Diet12}, P04:
  \citet{Pet04}, G12: \citet{Grier12} and est: estimated based on the
  relationship between black hole mass and near-IR virial product
  presented in \citet{L13}; (8) radius of the H$\beta$ broad-emission
  line region (in light-days); (9) reference for the H$\beta$ radius,
  where B09: \citet{Bentz09}, est: estimated based on the relationship
  between H$\beta$ radius and $1~\mu$m continuum luminosity presented
  in \citet{L13}, and the remainder as in column (7); (10) outer
  radius (in light-days) calculated from the {\it half} width of the
  flat top using the virial theorem; (11) total accretion disc
  luminosity; (12) blackbody temperature of the hot dust derived from
  near-IR spectral continuum fits; and (13) hot dust radius (in
  light-days) estimated from the total accretion disc luminosity in
  column (11) and the hot dust temperature in column (12) using
  eq. (1) of \citet{Mor12} for a silicate dust grain composition.}

\medskip

\parbox[]{17.5cm}{$^\star$ We measure a redshift of $z=0.145$ from
  narrow emission lines, instead of $z=0.142$ listed in NED.}

\end{table*}


The target selection, observational strategy, and data reduction
procedures for our original sample have been described in detail in
\citet{L08a}. In short, we obtained for 23 relatively nearby
($z\la0.3$) and bright ($J\la14$ mag) broad emission line AGN during
four observing runs contemporaneous (within two months) near-IR and
optical spectroscopy. The observations were carried out between 2004
May and 2007 January with a single object being typically observed
twice within this period. The near-IR spectra were obtained with the
SpeX spectrograph \citep{Ray03} at the NASA Infrared Telescope
Facility (IRTF), a 3 m telescope on Mauna Kea, Hawai'i. We chose the
short cross-dispersed mode (SXD, $0.8-2.4$ $\mu$m) and a slit of
$0.8\times15''$, which resulted in an average spectral resolution of
full width at half-maximum (FWHM) $\sim 400$~km s$^{-1}$.  In
\citet{L08a}, we presented the data of the first three epochs (2004
May, 2006 January, and 2006 June) and in \citet{L11a} we added the
fourth epoch (2007 January).

We obtained near-IR spectroscopy for an additional sample of nine
broad emission line AGN in order to improve the calibration of the
near-IR relationship for estimating black hole masses presented in
\citet{L11b}. Details on the target selection, observational strategy,
and data reduction procedures can be found in \citet{L13}. In short,
we observed in queue mode at the {\it Gemini} North observatory, an
8.1~m telescope on Mauna Kea, Hawai'i, with the Gemini Near-Infrared
Spectrograph \citep[GNIRS;][]{gnirs} in its Science Verification (SV)
phase and in semester 2011B. We used the cross-dispersed mode with the
short camera ($0.9-2.5$~$\mu$m) and a slit of $0.675 \times7''$. This
set-up gives an average spectral resolution of FWHM $\sim
400$~km~s$^{-1}$, similar to that of the IRTF near-IR spectra.

All spectra have relatively high continuum signal-to-noise (S/N)
ratios ($\ga 100$), which allows us to study even the weakest lines in
the Paschen series and to interpret the profiles of the strongest
Paschen lines, Pa$\alpha$ and Pa$\beta$, with high confidence. In the
case of the IRTF data, we have used for our main analysis only the
highest-quality spectrum. Furthermore, we have excluded the source
Mrk~590, since it was found to be in a very low AGN state with weak
and noisy broad emission lines. Our total sample consists of 31
sources, for which we list in Table \ref{sample} the relevant
measurements for the broad components of the Pa$\alpha$ or Pa$\beta$
emission lines, whichever had the higher S/N ratio. In the following,
we assume that these broad components are representative of the
hydrogen BELR and refer to them as 'template profiles'.

\begin{figure}
\centerline{
\includegraphics[clip=true, bb=24 390 589 717, scale=0.42]{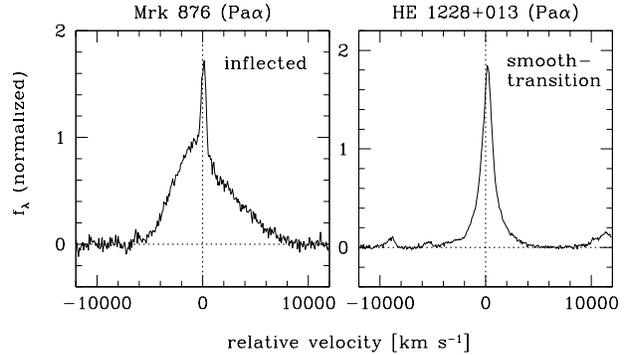}
}
\caption{\label{inflest} Observed Pa$\alpha$ emission lines in
  velocity space relative to the expected rest-frame wavelength. The
  separation between the broad and narrow line components is obvious
  when the profile has an inflection (left panel), but not so when the
  emission lines are relatively narrow and the broad and narrow line
  components smoothly transition into each other (right panel).}
\end{figure}

The profiles of the total observed Pa$\alpha$ or Pa$\beta$ emission
lines of our sources fall roughly into two subclasses depending on the
transition between the broad and narrow line components. The
separation between the two components is obvious when the total
profile has an inflection (as exemplified in Fig. \ref{inflest}, left
panel), but not so when the emission lines are relatively narrow and
the broad and narrow line components smoothly transition into each
other (Fig. \ref{inflest}, right panel). In the following we refer to
these two subclasses of sources as 'inflected' and 'smooth-transition
sources', respectively. In our sample, the total observed profiles of
14/31 sources is inflected and that of 13/31 sources is of the
smooth-transition type. A further four sources (PG~0804$+$761,
PG~0844$+$349, Ark~120 and NGC~3516) clearly lack a Paschen narrow
line component, since their profiles have a broad top. For the
analysis in Section \ref{dust} we have grouped these four sources with
the inflected ones. Fig. \ref{BLRinfl} and \ref{BLRest} show the
template profiles separately for the inflected and smooth-transition
sources. The three inflected sources 3C~120, NGC~7469 and NGC~3227
have been included in Fig. \ref{BLRest} rather than in
Fig. \ref{BLRinfl} since they have relatively narrow broad emission
lines. In the next section, we discuss our novel method to separate
the broad and narrow emission line components in smooth-transition
sources.

\section{Absent narrow components} \label{narrow}

\begin{figure*}
\centerline{
\includegraphics[clip=true, bb=24 175 589 717, scale=0.95]{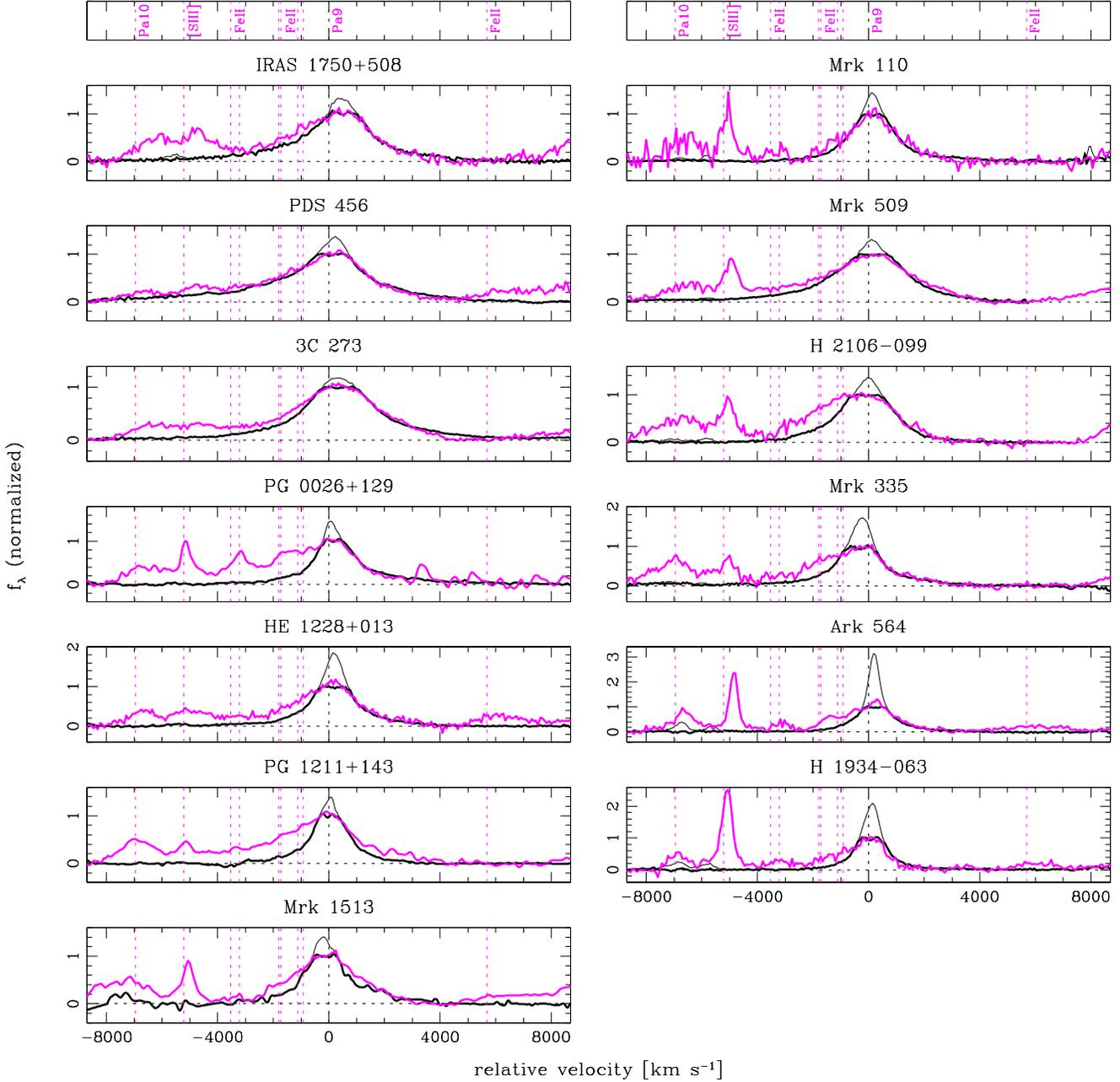}
}
\caption{\label{Pa9} Observed profiles of the Pa9 and Pa10 emission
  lines (magenta) for the smooth-transition sources. The Pa9 profiles
  are compared to the template profiles (Pa$\alpha$ or Pa$\beta$,
  black) in velocity space relative to the expected rest-frame
  wavelength. The absent Pa9 narrow component reveals a flat-topped
  broad emission line profile similar to that resulting for the
  template profiles (thick lines) after removing the largest possible
  flux contribution from the narrow emission line region (thin
  lines).}
\end{figure*}

\begin{figure*}
\centerline{
\includegraphics[clip=true, bb=24 300 590 715, scale=0.9]{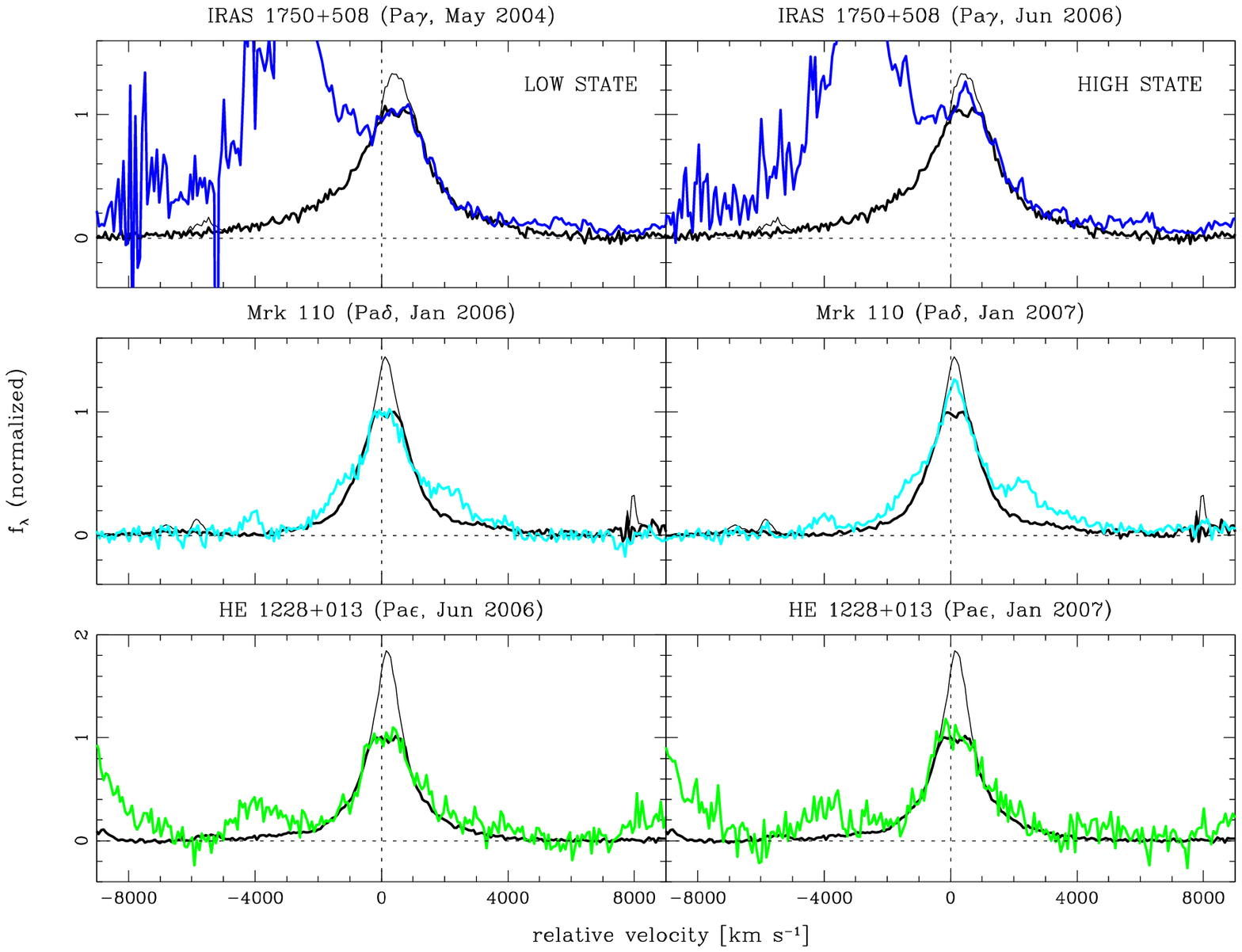}
}
\caption{\label{PaDPaE} Observed profiles of the Pa$\gamma$ (blue),
  Pa$\delta$ (cyan) and Pa$\epsilon$ (green) emission lines for the
  smooth-transition sources IRAS~1750$+$508, Mrk~110 and
  HE~1228$+$013, respectively, compared to the template profiles
  (Pa$\alpha$ or Pa$\beta$, black) in velocity space relative to the
  expected rest-frame wavelength. When the source is in a low flux
  state (left-hand panels), the absent Pa$\gamma$, Pa$\delta$ and
  Pa$\epsilon$ narrow components reveal a flat-topped broad emission
  line profile. This profile is similar to that resulting for the
  template profiles (thick lines) after removing the largest possible
  flux contribution from the narrow emission line region (thin
  lines). When the source is in a high flux state (right-hand panels),
  the narrow component of the Pa$\epsilon$ emission line is absent in
  HE~1228$+$013, whereas the narrow components of the Pa$\gamma$ and
  Pa$\delta$ emission lines are present in IRAS~1750$+$508 and
  Mrk~110, respectively.}
\end{figure*}

Since the first optical spectroscopy of AGN it was realised that not
all broad emission line profiles are peaked but that some have a very
broad, flat top \citep[e.g.,][]{Ost75, Ost76}. The narrow emission
line component would then sit on top of these very broad profiles or
be outright absent. In these AGN, it is evident that the BELR has an
outer boundary and that it does not kinematically merge into the
NELR. However, it is not immediately clear what the intrinsic broad
emission line profile is when no flat top is observed and the total
profile resembles rather that of a narrow emission line with a broad
bottom and extended wings. In this section, we show that also in these
AGN the broad emission line profiles are intrinsically flat-topped.

A crucial step towards isolating the intrinsic broad emission line
profile in AGN is the subtraction of the narrow line component. This
task is straightforward when the transition between the broad and
narrow emission line components is obvious because there is a clear
inflection in the profile, which is generally the case for broad
components with FWHM a factor of $\ga 5$ larger than those of the
narrow components \citep{L08a}. The profiles of roughly half our
sample fall in this category (exemplified in Fig. \ref{inflest}, left
panel). When the broad emission lines are relatively narrow and the
broad and narrow components smoothly transition into each other (see
Fig. \ref{inflest}, right panel), two different approaches have been
adopted so far to isolate the intrinsic broad emission line
profile. The narrow component is assumed to be a Gaussian with FWHM
the same as that of the forbidden narrow emission line \OIII~$\lambda
5007$, which is taken to be representative of the NELR, and the flux
of the Gaussian is either scaled to that of \OIII~$\lambda 5007$
assuming Case B conditions \citep[e.g.][]{Marz96, Marz03} or is fit
for by modeling the entire emission line profile with the sum of two
or three Gaussians \citep[e.g.][]{Zamfir10, Jin12}. Both these methods
yield an intrinsic broad emission line profile that is {\it
  peaked}. In \citet{L08a}, we have taken yet a different approach and
have fit in these cases a Gaussian with FWHM equal to that of
\OIII~$\lambda 5007$ to the {\it entire} top part of the total
emission line profile. For the additional sample presented in
\citet{L13}, for which we did not have contemporaneous optical
spectroscopy, we have used instead the FWHM of the near-IR narrow
emission line \SIII~$\lambda 9531$. Our new method assumes that the
NELR and BELR are kinematically distinct and by scaling the flux of
the Gaussian to the entire upper part of the emission line profile it
subtracts the largest possible flux contribution from the
NELR. Contrary to the other two methods, it yields an intrinsic broad
emission line profile that is flat-topped, similar to what is observed
in the inflected sources.

In support of our method, we present here a new finding; in {\it all}
smooth-transition sources the narrow components of the higher-order
Paschen emission lines, such as, e.g., Pa9 and Pa10 (Fig. \ref{Pa9}),
are absent. In some of these AGN, in particular when they are in a low
flux state, we find absent narrow components even earlier in the
Paschen series, e.g., Pa$\gamma$, Pa$\delta$ and Pa$\epsilon$ for the
sources IRAS~1750$+$508, Mrk~110 and HE~1228$+$013, respectively
(Fig. \ref{PaDPaE}). The absent narrow components reveal the intrinsic
broad-line profile of the smooth-transition sources, which is observed
to be flat-topped and similar to both the profile that our method
yields for the Pa$\alpha$ and Pa$\beta$ broad components and the
Paschen broad-line profile of the inflected sources. The most likely
reason for the absent narrow components of the higher-order Paschen
emission lines is the much steeper flux decrement within the series
for conditions prevelant in the NELR (i.e. for optically thin Paschen
line emission) than for those typical of the BELR (i.e. for optically
thick Paschen line emission). We note that the Pa$\epsilon$ broad
emission line is heavily blended with the strong narrow emission line
\SIII~$\lambda 9531$ and, therefore, we cannot exclude that the
Pa$\epsilon$ narrow component is absent also in other sources. The
source HE~1228$+$013 is the only one in our sample that lacks
\SIII~$\lambda 9531$~emission. Furthermore, based on the observed
flat-topped Pa9 broad emission lines, we have now estimated the
contribution of the narrow components to the template profiles also
for the sources IRAS~1750$+$508 and PDS~456, which were left
uncorrected by \citet{L08a}.

\section{A dust-limited BELR} \label{dust}

An intrinsically flat-topped broad emission line profile in {\it all}
AGN indicates that the BELR generally has an outer radius. Next, we
investigate if its value is set by the presence of dust, i.e. if the
BELR is dust- rather than radiation-bounded. As \citet{Netzer93} have
shown, when dust is present, the broad line emission diminishes
sharply and narrow line emission is expected at much larger radii.

Assuming the gas dynamics are dominated by gravity, we have used the
virial theorem $M_{\rm BH} = f R \Delta V^2/G$, where $M_{\rm BH}$ is
the black hole mass, $R$ is the radial distance of the broad line gas,
$\Delta V$ is the gas velocity, $G$ is the gravitational constant and
$f$ is the geometrical correction factor, to calculate the outer
radius of the BELR. Half of the observed full width of the flat top
readily gives $\Delta V$. We have measured the width of the flat top
in the template profiles as indicated by the two vertical red dashed
lines in Figs. \ref{BLRinfl} and \ref{BLRest}. For the large majority
of our sample (24/31 sources) we have black hole masses determined by
optical reverberation mapping campaigns and for the remainder we have
estimated this value based on our recently calibrated relationship
between the black hole mass and the near-IR virial product
\citep{L13}. The geometrical correction factor $f=1$ in our case,
since, contrary to the line dispersion or FWHM of the broad line, the
outer radius and so the flat top width is independent of optical depth
effects \citep{Kor04}. The results are listed in Table
\ref{sample}. We obtain BELR outer radii in the large range of
\mbox{$R_{\rm out} \sim 26$~lt-days} $- 153$~lt-yrs.

\begin{figure}
\centerline{
\includegraphics[scale=0.42]{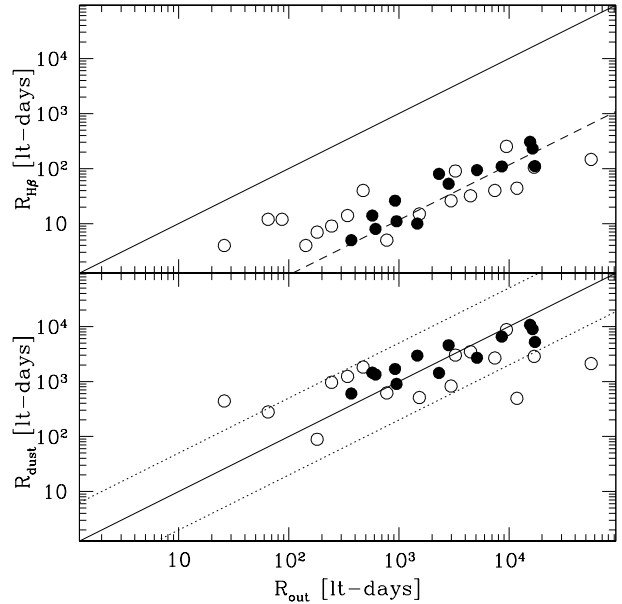}
}
\caption{\label{Rout} The reverberation radius of the H$\beta$ broad
  line region (top panel) and the hot dust radius (calculated for a
  silicate dust grain composition; bottom panel) versus the BELR outer
  radius. Sources with template profiles (Pa$\alpha$ or Pa$\beta$)
  that have an obvious and an estimated transition between their broad
  and narrow components are shown as open and filled circles,
  respectively. In both panels we show the line of equality (solid
  line) with the dashed line indicating the mean for the entire sample
  of $R_{\rm out}/R_{\rm H\beta}=86$ (top panel) and the dotted lines
  showing the range of a factor of 5 around the line of equality
  (bottom panel).}
\end{figure}

In Fig. \ref{Rout} (top panel), we first compare the resulting BELR
outer radii to the radii of the H$\beta$ broad line region obtained
from reverberation mapping campaigns. For the seven sources without
reverberation results we have estimated this value based on our
recently calibrated relationship between the H$\beta$ radius and the
$1~\mu$m continuum luminosity \citep{L13}. We expect the H$\beta$
broad line gas at the largest radii to respond the strongest to
changes in the ionising continuum flux, but at the same time emission
line gas at smaller radii can respond more rapidly and more coherently
than gas at larger radii, which will bias observed reverberation time
lags towards smaller values \citep{Kor00, Kor04}. Therefore, it is
difficult to theoretically predict the expected ratio between the BELR
outer radius and the reverberation radius and its empirical value can
put important constraints on the covering fraction and radial density
distribution of the line emitting gas. For our entire sample, we
obtain a relatively large range of \mbox{$R_{\rm out}/R_{\rm H\beta}
  \sim 5 - 380$}, with a distribution mean of $\langle R_{\rm
  out}/R_{\rm H\beta} \rangle = 86\pm15$ (dashed line) and a median
value of $R_{\rm out}/R_{\rm H\beta}=54$. The sources NGC~5548 and
PG~0804$+$761 have the smallest and largest radius ratio,
respectively. The large spread in values is mainly due to the
inflected sources. If we consider the inflected and smooth-transition
sources separately, the distribution means are $\langle R_{\rm
  out}/R_{\rm H\beta} \rangle = 96\pm25$ and $73\pm11$, with
corresponding median values of $R_{\rm out}/R_{\rm H\beta}=37$ and 71,
respectively. The much larger (a factor of $\sim 2$) median value for
the smooth-transition sources indicates that the gas emissivity, which
strongly depends on the number density, in these sources does not drop
as rapidly with radius as in the inflected sources, suggesting that
they have much lower ionisation parameters. We will return to this
point later in the section.

\begin{figure}
\centerline{
\includegraphics[scale=0.42]{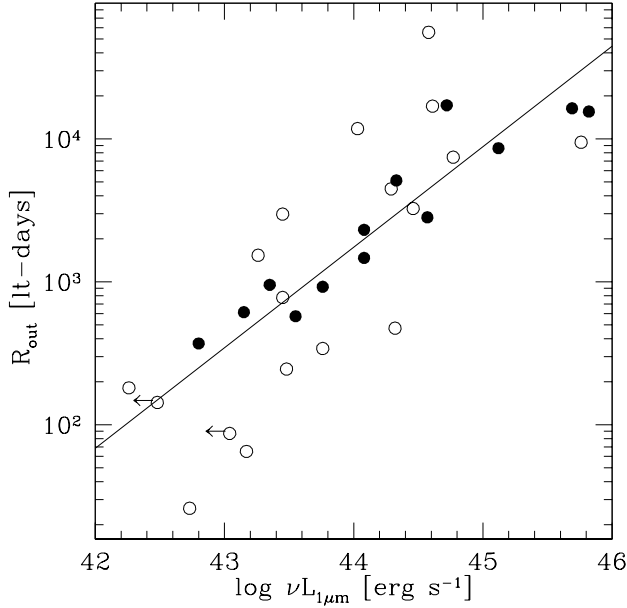}
}
\caption{\label{outradius} The outer radius (in light-days) versus the
  integrated $1~\mu$m continuum luminosity. Symbols are as in
  Fig. \ref{Rout}. The two sources with a continuum SED dominated by
  host galaxy emission are plotted as upper limits. The solid line
  indicates the observed correlation.}
\end{figure}

If the BELR is indeed dust-limited, we expect its emission to cease at
a certain incident continuum photon flux, which should roughly
coincide with that required for dust sublimation. On the other hand,
the dust that possibly limits the BELR is most likely the same as the
hot dust usually observed in the spectral energy distributions (SEDs)
of broad emission line AGN, and so we expect a rough coincidence
between the BELR outer radius and the average hot dust radius. In the
following we search for these two lines of evidence. First, in order
to assess if there is a preferred ionisation flux that limits the
BELR, we have plotted in Fig. \ref{outradius} the radius-luminosity
($R$-$L$) relationship between the BELR outer radius and the
integrated 1~$\mu$m continuum luminosity. As we have shown in
\citet{L11a}, the latter is dominated by the accretion disc spectrum,
which is believed to be the main source of ionising radiation in
AGN. The continuum SEDs of two of our sources (NGC~3516 and NGC~3227)
are dominated by host galaxy emission and, therefore, their observed
1~$\mu$m continuum luminosities represent only an upper limit to the
AGN luminosity. Fig. \ref{outradius} shows that there is a strong
correlation between the BELR outer radius and the ionising
luminosity. A least-squares fit (excluding the two upper limits) gives
a correlation slope of $0.7\pm0.1$, which is consistent (within
$2\sigma$) with the value of 0.5 expected from simple photoionisation
arguments. Therefore, it appears that the BELR in all AGN ceases at a
similar ionising {\it flux}. Fig. \ref{outradius} further supports our
method for removing the narrow line components in smooth-transition
sources. It shows that both the inflected and smooth-transition
sources follow the same $R$-$L$ relationship, which means that we have
not artificially introduced a broad, flat top in the latter. If that
had been the case, for a given continuum luminosity, we would expect
to find the smooth-transition sources at much lower BELR outer radii
than those of the inflected sources.

Next, we have estimated the hot dust radius in our sources using the
relationship between bolometric luminosity and dust sublimation
temperature given by \citet{Mor12}. We have assumed that the dust
sublimation temperature corresponds to the hot dust temperature and
have derived the latter from blackbody fits to the near-IR spectral
continuum as described in \citep{L11a}. For our enlarged sample, the
average hot dust temperature is $T_{\rm dust} = 1347 \pm 20$~K,
similar to our result in \citet{L11a}, which indicates a silicate dust
grain composition. Therefore, we have used eq. (1) of
\citet{Mor12}. We have approximated the bolometric luminosity with the
total accretion disc luminosity, which we have obtained from spectral
continuum fits as discussed in \citet{L11a}. Our results are listed in
Table \ref{sample}. Note that for two sources, namely, NGC~3516 and
NGC~3227, we do not have an estimate of the hot dust radius since
their near-IR spectra were found to be strongly dominated by host
galaxy starlight \citep{L11a, L13}. In Fig. \ref{Rout} (bottom panel),
we compare the hot dust radii to the BELR outer radii. The
distribution of the ratio between the two radii has a mean consistent
with one ($\langle R_{\rm dust}/R_{\rm out} \rangle = 1.9\pm0.6$) and
a median value of $R_{\rm dust}/R_{\rm out}=0.8$. As previously in
Fig. \ref{Rout} (top panel), the inflected sources show a much larger
range in values than the smooth-transition sources. If we consider
only the latter sources, we get a mean of $\langle R_{\rm dust}/R_{\rm
  out} \rangle = 1.2\pm0.2$ and a median value of $R_{\rm dust}/R_{\rm
  out}=0.95$.

The largest source of uncertainty in our estimation of both the BELR
outer radius and the hot dust radius is the black hole mass, which
enters through the virial theorem in the former and the accretion disc
fits in the latter. The virial black hole mass is known only within a
factor of $\sim 2-3$ \citep[e.g.][]{Graham11}, which translates to an
uncertainty in the ratio between the two radii of a factor of $\sim
3-5$. Another unknown is the chemical composition of the hot dust. If
we assume instead graphite dust grains and use eq. (2) of
\citet{Mor12}, the estimated hot dust radii reduce by a factor of 1.5.
Given these uncertainties, it is remarkable that for the large
majority of our sample the BELR outer radius and the hot dust radius
are consistent with each other within a factor of $\la 5$. Only four
sources are the exception, namely, PG~0804$+$761, PG~1307$+$085,
3C~390.3, and NGC~4593, which are all of the inflected
type. Therefore, it appears that in general the BELR is dust-limited.

We can compare our results also to the lag times measured from near-IR
dust reverberation programmes. \citet{Kosh09} found that the hot dust
radius in the source NGC~4151 changed between 30 and 70 days, which is
a factor of $\sim 3-6$ lower than our BELR outer radius of 181
lt-days, but consistent with our estimated hot dust radius of 89
lt-days. \citet{Sug06} found a lag time of $\sim 50$ days for the
source NGC~5548, consistent with our value of the BELR outer radius of
65 lt-days, but a a factor of $\sim 6$ lower than our estimated hot
dust radius of 279 lt-days. On the other hand, their lag times for the
sources NGC~3227 ($\sim 20$ days) and NGC~7469 (65$-$87 days) are
considerably lower (by a factor of $\sim 7$ and $\sim 9-12$,
respectively) than our BELR outer radii. For the source NGC~7469,
their lag times are also much lower (by a factor of $\sim 7-10$) than
our estimated hot dust radius of 620 lt-days, whereas we have no
estimate of the hot dust radius for the source NGC~3227 since both its
near-IR and optical spectra were found to be strongly dominated by the
host galaxy starlight \citep{L11a}. The fact that lag times measured
by dust reverberation programmes are systematically smaller than dust
radii estimated from the AGN bolometric luminosity and the dust
sublimation temperature has been noted previously \citep{Okn01,
  Kish07, Nen08a}. \citet{Kaw10, Kaw11} ascribe this to a dust
geometry that is bowl-shaped rather than spherical due to the
anisotropy of the accretion disc emission. Such a geometry would place
the dust further away from the continuum source but nearer to the
observer for pole-on views, thus reducing the observed dust
reverberation times relative to the dust radius estimated from the
continuum radiation.

\citet{Netzer93} showed that dust strongly reduces the emissivity of
the broad line gas. Then, assuming that dust is present at all
distances outside the BELR, emission from the NELR will appear only at
much larger radii and an obvious, large gap in velocity field between
the BELR and NELR is expected. This is observed in the inflected
sources, but it is not immediately clear how the smooth-transition
sources can fit in this scenario. One possibility is that the
ionisation parameter is smaller in the latter than in the former
sources, as already suggested by our results based on Fig. \ref{Rout}
(upper panel). Since the effects of dust on the gas emissivity due to
its absorption of ionising photons and destruction of line photons are
stronger the higher the ionisation parameter, a relatively small
ionisation parameter in the smooth-transition sources would mean that
the dust barely affects the line emissivity thus producing only a
small, unperceptible velocity gap between the BELR and NELR.

The ionisation parameter, which is defined as \mbox{$U=\Phi/(n\cdot
  c)$}, where $\Phi$ is the ionising flux, $n$ is the number density
and $c$ is the speed of light, is a measure of the number of photons
available to ionise versus the number of atoms available to be
ionised. Since the ionising luminosity is on average similar for the
inflected and smooth-transition sources ($\langle \log \nu L_{\rm tot}
\rangle = 45.45\pm0.23$ and $46.22\pm0.22$, respectively), a lower
ionisation parameter in the latter would imply a line emitting gas of
a higher number density. Such high-density gas is expected to produce
copious emission from low-ionisation species such as, e.g., \OI,
\FeII, and~\CaII, and this even at large distances from the ionising
source. Indeed, in \citet{L08a}, we showed that whereas the inflected
sources have only broad \OI~emission, the \OI~$\lambda 8446$ and
\OI~$1.1287$~$\mu$m emission line profiles of the smooth-transition
sources show both a broad and a narrow component.

\section{The geometry and kinematics}
 
All current BELR models assume that the gas motion is dominated by the
gravitational potential of the central black hole, i.e. that the gas
is virialised. The differences between them lie mainly in the geometry
of the gas distribution and the presence or absence of a radial
component. In this section, we briefly discuss under which assumptions
the most popular BELR models are able to produce a broad line profile
with a flat top. In the following, we consider (i) a spherical gas
distribution in orbital motion; (ii) a pure Keplerian disc; and (iii)
a Keplerian disc with a wind.

A flat-topped broad line profile is produced by a spherical gas
distribution in orbital motion (or in gravitational infall) if a sharp
outer boundary rather than an indefinetely large outer radius is
assumed. This has been noted by several authors who have calculated
detailed broad line profiles from such models \citep[e.g.][]{Capr80,
  Rob90, Rob95, Corb97b, Kor04}. However, the assumption of a sharp
outer boundary was always considered to be artificial, since most
observed broad emission line profiles were peaked. In spherical
models, a flat-topped profile is expected for ratios between the BELR
outer and inner radius of $R_{\rm out}/R_{\rm in} \la 200$.

A pure Keplerian disc produces for most orientations and radial
extents of the disc a double-horned broad emission line profile
\citep[e.g.][]{Horne86, Chen89, Dum90}. The two horns are separated by
velocities of \mbox{$\ga 600$~km~s$^{-1}$}, and, therefore, a flat top
is generally not expected. The only exception are profiles predicted
for relatively small angles between the observer and the disc rotation
axis of $\la 15^{\circ}$, for which the two horns almost merge
together. However, in this case the emission line width is also
considerably reduced. A bowl-shaped geometry, as recently proposed by
\citet{Goad12}, produces broad emission line profiles largely similar
to those of a pure Keplerian disc. We note that the Balmer broad
emission lines H$\alpha$ and H$\beta$ of the source 3C~390.3, which is
included in our sample, have been previously modelled with a disc
\citep{Perez88, Zheng91, Era94}. We observe a Pa$\alpha$ profile for
3C~390.3 that is clearly flat-topped, but a strong wing extending to
very large velocities is also apparent.

\citet{Murray95} have shown that if a wind is (radiatively) launched
from a disc in Keplerian rotation, the expected broad emission line
profile changes from double-horned to single-peaked, since the
substantial radial velocity shear in the wind considerably increases
the optical depth and so more line photons escape along directions
with low projected velocity. In this model, the emission lines arise
in a thin layer where the wind emerges from the disc, which means that
the motion of the line emitting gas is largely dominated by
gravity. \citet{Murray97} show that a flat-topped broad emission line
profile is produced in the disc wind model if relatively small ratios
between the BELR outer and inner radius are assumed ($R_{\rm
  out}/R_{\rm in} \la 100$). However, the smaller the BELR outer
radius, the smaller also the expected equivalent width of the emission
line. The accretion disc wind model has recently been developed
further by \citet{Flohic12} and \citet{Chajet13} for the low- and
high-ionisation emission lines, respectively. In particular the models
of \citet{Flohic12} for the Balmer emission lines show that the width
of the flat top also depends on the inclination of the disc; the
broadest flat tops are expected for edge-on views.

\section{Summary and conclusions}

We have used the Paschen hydrogen emission lines of a sample of 31
type 1 AGN to show that the BELR generally has an outer boundary. The
intrinsic broad emission line profile is flat-topped, which means that
the BELR and NELR are kinematically separate. Our near-IR spectra,
which were obtained in cross-dispersed mode at the IRTF 3~m and Gemini
North 8~m observatories, have both a large wavelength coverage and a
high continuum S/N ratio ($\ga 100$). This allowed us to study even
the weakest lines in the Paschen series and to interpret the unblended
profiles of the strongest Paschen lines, Pa$\alpha$ and Pa$\beta$,
with high confidence. Our main results can be summarised as follows.

\medskip

(i) The higher-order Paschen lines (in particular the Pa9 and Pa10
lines) provide observational evidence that the BELR has an outer
radius. Their narrow components are absent in {\it all} sources,
revealing a broad emission line profile that is intrinsically
flat-topped. This result indicates that the BELR and NELR are
kinematically separate and is most relevant for AGN with relatively
narrow broad emission lines, for which the transition between the
broad and narrow line components is not obvious.

(ii) We have calculated the BELR outer radius from the {\it half}
width of the flat top in the Pa$\alpha$ or Pa$\beta$ profiles using
the virial theorem. The resulting values follow a radius-luminosity
relationship with a logarithmic slope of $0.7\pm0.1$, which is
consistent (within $2\sigma$) with the value of 0.5 expected from
simple photoionisation arguments. This indicates that the BELR in all
AGN ceases at a similar ionsing flux.

(iii) We have estimated the dust sublimation radius from the AGN
bolometric luminosity (approximated with the total accretion disc
luminosity) and the hot dust temperature and find that it is on
average similar to the BELR outer radius. The distribution of the
ratios between the two radii has a mean of $\langle R_{\rm
  dust}/R_{\rm out} \rangle = 1.9\pm0.6$. Sources with an obvious
transition between their broad and narrow line components show a much
larger range in values than those with relatively narrow broad
emission lines. If we consider only the latter sources, the mean moves
closer to one ($\langle R_{\rm dust}/R_{\rm out} \rangle =
1.2\pm0.2$). Although a dust-bounded BELR was already proposed early
on \citep{Netzer93}, we present a firm observational evidence for it.

(iv) The observation of an intrinsically flat-topped broad emission
line profile can constrain the geometry and kinematics of the
BELR. Spherical models with orbital motion or infall produce a flat
top if a sharp outer boundary rather than an indefinetely large outer
radius is assumed. This is also the case for the accretion disc wind
model of \citet{Murray95}. The required ratios between the BELR outer
and inner radius in these models are $\la 100 - 200$. On the other
hand, a pure Keplerian disc produces for most orientations and radial
extents of the disc a double-horned broad emission line profile.

\section*{Acknowledgments}

We thank Kirk Korista and Brad Peterson for their comments on an
earlier version of this manuscript. H. L. acknowledges financial
support by the European Union through the COFUND scheme. This work is
partly based on observations obtained at the {\it Gemini} Observatory,
which is operated by the Association of Universities for Research in
Astronomy, Inc., under a cooperative agreement with the NSF on behalf
of the Gemini partnership: the National Science Foundation (United
States), the Science and Technology Facilities Council (United
Kingdom), the National Research Council (Canada), CONICYT (Chile), the
Australian Research Council (Australia), Minist\'erio da Ci\^encia,
Tecnologia e Inova\c{c}\~ao (Brazil) and Ministerio de Ciencia,
Tecnolog\'ia e Innovaci\'on Productiva (Argentina).

\bibliography{/Users/herminelandt/references}

\appendix

\section{Paschen broad emission line profiles}

\begin{figure*}
\centerline{
\includegraphics[clip=true, bb=24 170 589 717, scale=0.95]{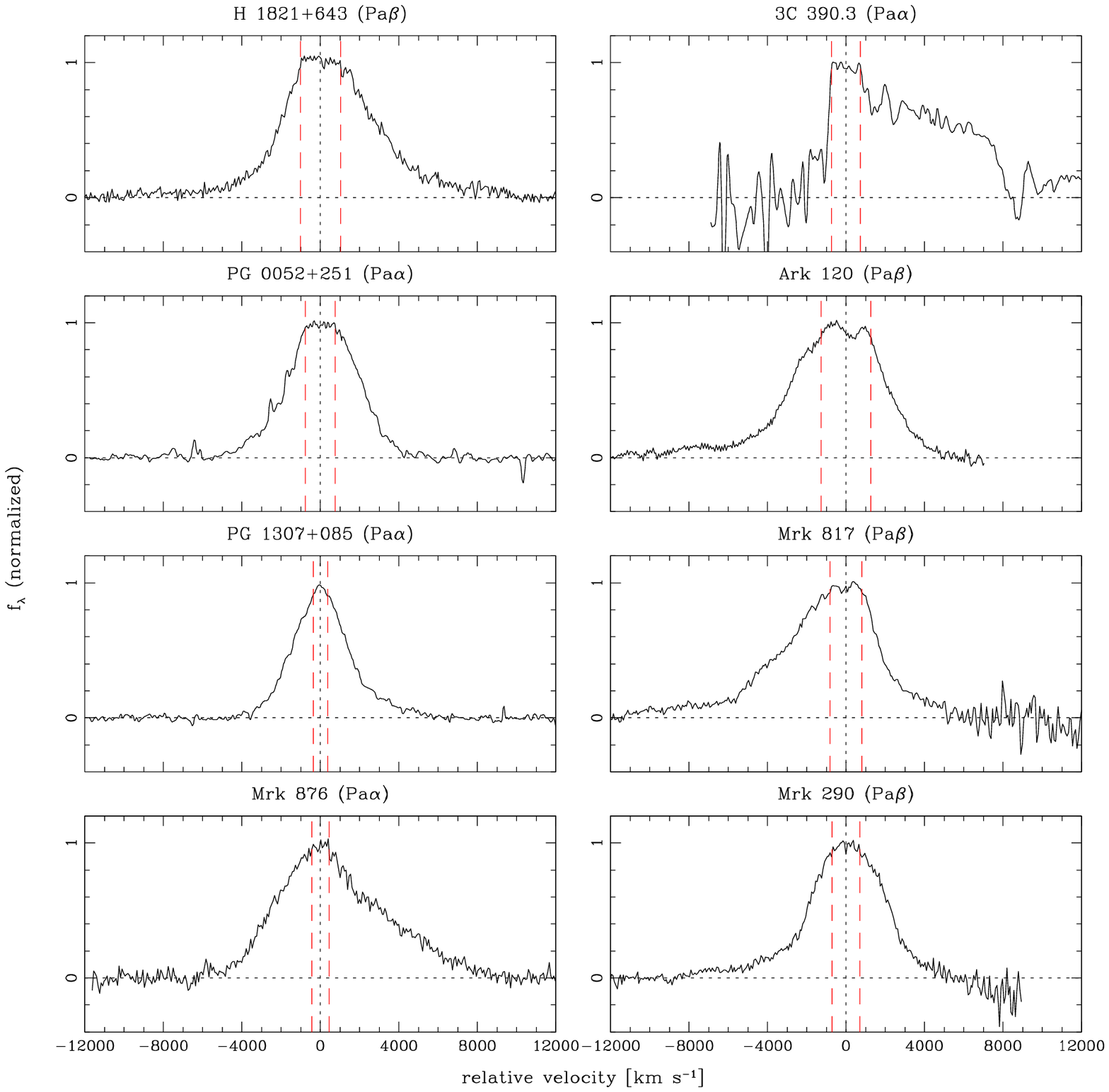}
}
\caption{\label{BLRinfl} Profiles of the Pa$\alpha$ or Pa$\beta$ broad
  components (whichever had the higher S/N ratio) in velocity space
  for the inflected sources, i.e. sources with an obvious transition
  between the broad and narrow emission line components. The line
  profiles have been normalised to the same intensity of the flat top
  and shifted such that the flat top centre (vertical black dotted
  line) is at zero velocity. The two vertical red dashed lines mark
  the full width of the broad, flat top.}
\end{figure*}

\setcounter{figure}{1}
\begin{figure*}
\centerline{
\includegraphics[clip=true, bb=24 300 589 717, scale=0.95]{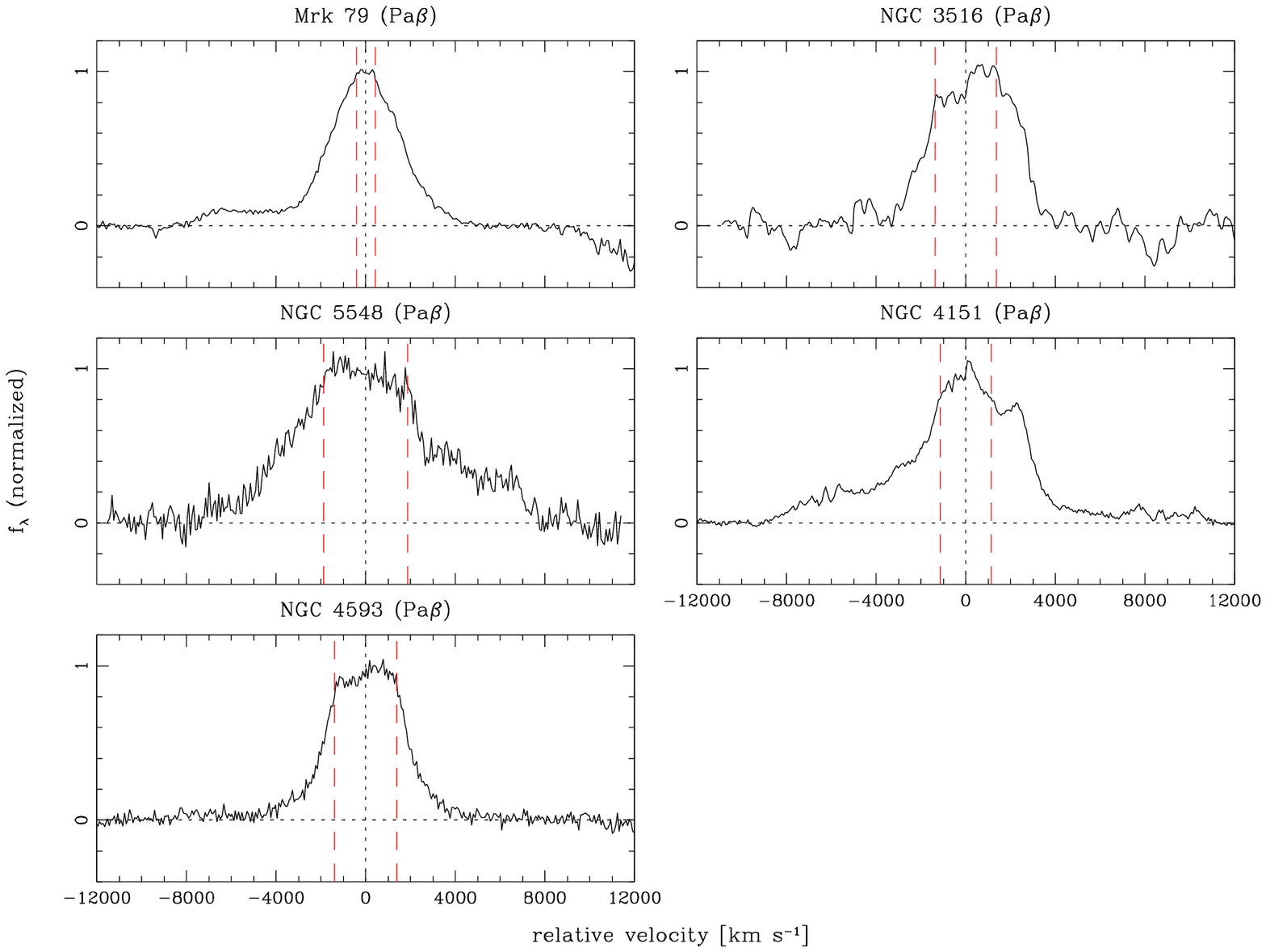}
}
\contcaption{}
\end{figure*}

\begin{figure*}
\centerline{
\includegraphics[clip=true, bb=24 170 589 717, scale=0.95]{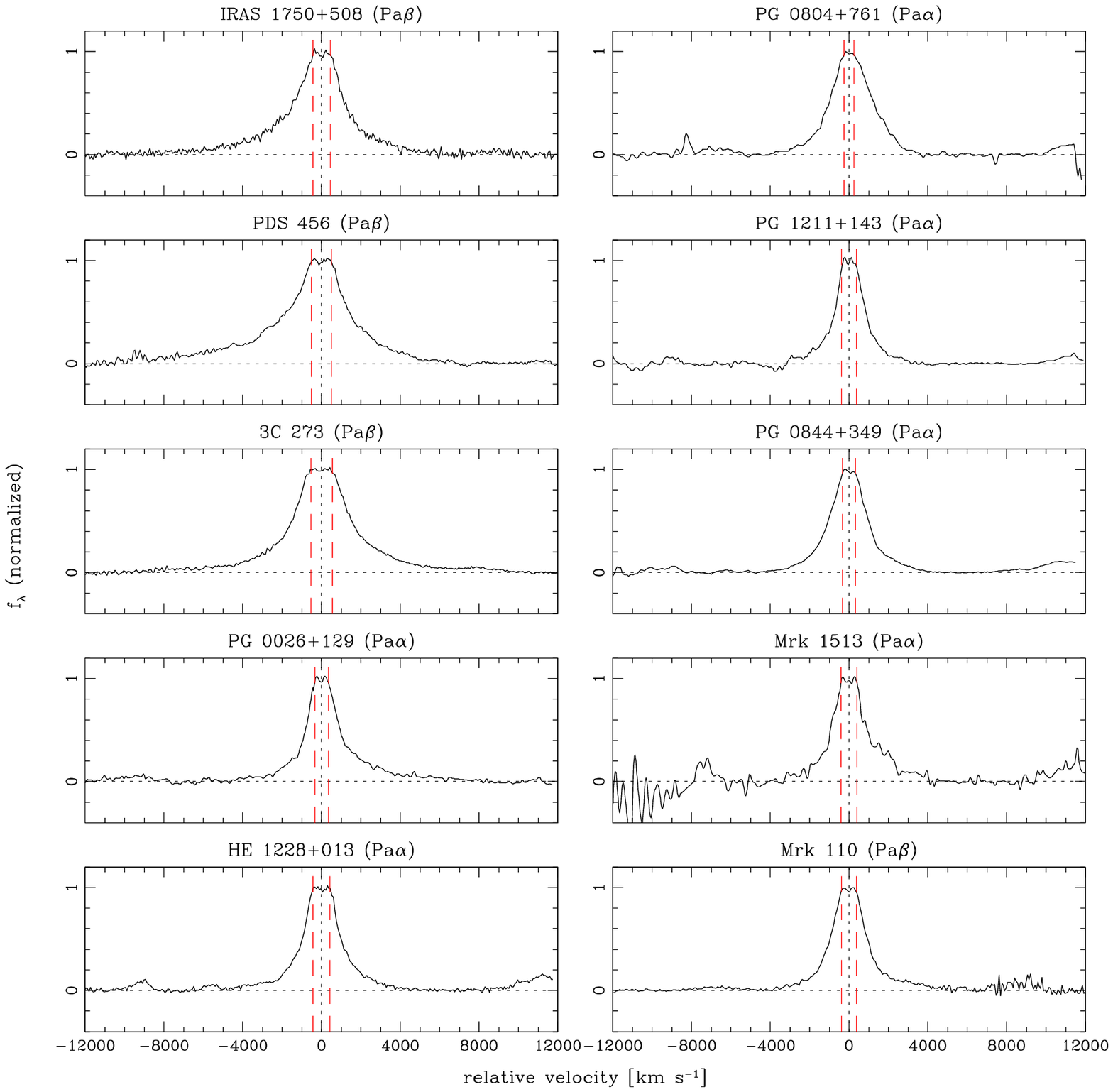}
}
\caption{\label{BLRest} Same as in Fig. \ref{BLRinfl} for the
  smooth-transition sources, i.e. sources with an estimated transition
  between the broad and narrow emission line components. The three
  inflected sources 3C~120, NGC~7469 and NGC~3227 with their
  relatively narrow broad components have also been included.}
\end{figure*}

\setcounter{figure}{2}
\begin{figure*}
\centerline{
\includegraphics[clip=true, bb=24 275 589 717, scale=0.95]{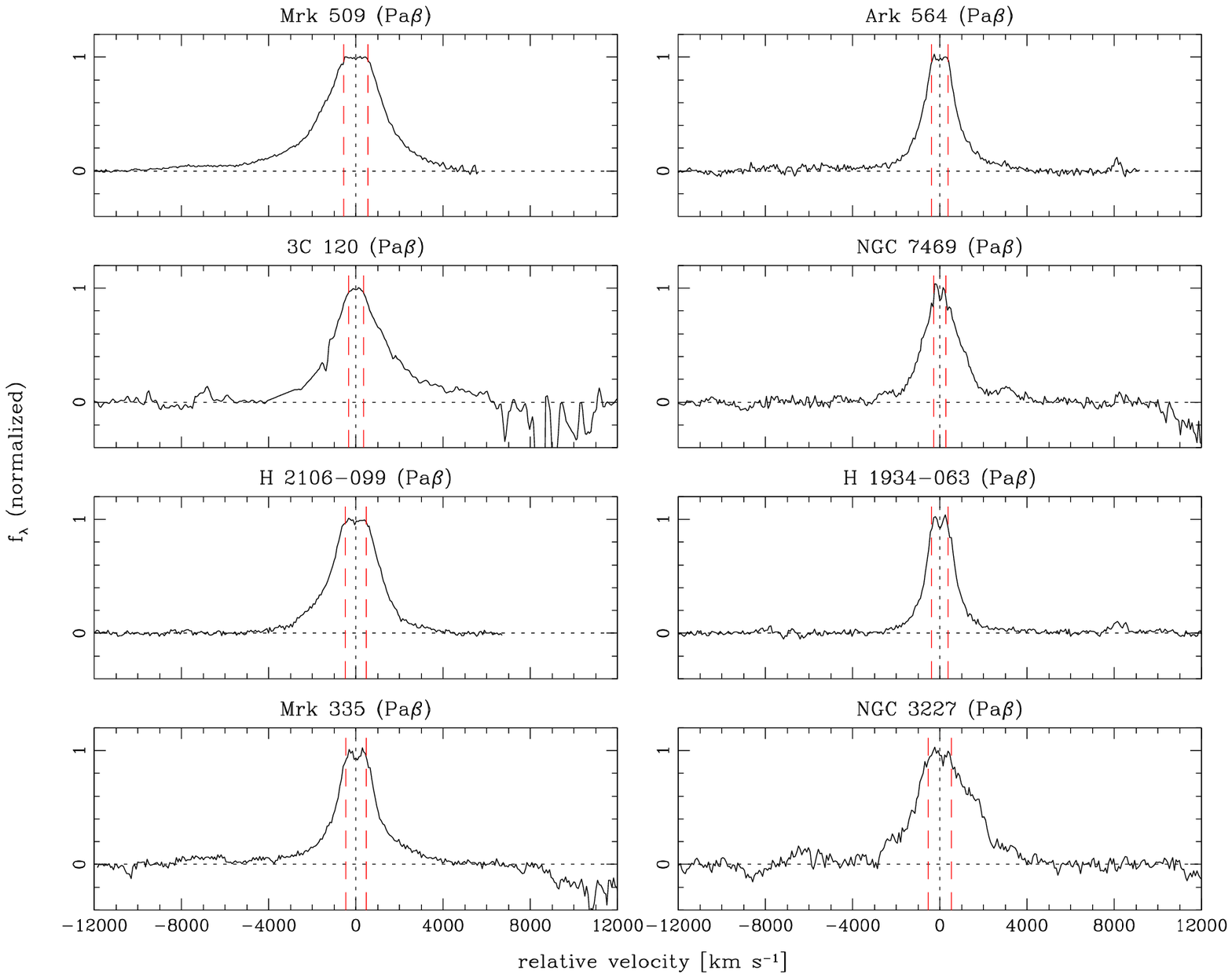}
}
\contcaption{}
\end{figure*}


\bsp
\label{lastpage}

\end{document}